\documentclass[twocolumn]{revtex4}
\usepackage{amssymb}

\input{tcilatex}

\begin{document}

\title{Femtosecond quasiparticle relaxation dynamics and probe polarization
anisotropy in YSr$_{x}$Ba$_{2-x}$Cu$_{4}$O$_{8}$ ($x=0,0.4$).}
\author{D.Dvorsek$^{1}$, V.V.Kabanov$^{1}$, J.Demsar$^{1,2}$, J.Karpinski$^{3}$,
S.M.Kazakov$^{3}$ and D.Mihailovic$^{1}$}
\affiliation{$^{1}$Jozef Stefan Institute, Jamova 39, 1000 Ljubljana, Slovenia\\
$^{2}$Los Alamos National Laboratory, Los Alamos NM 87545, USA\\
$^{3}$Institut fur Festkorperphysik, ETH Zurich, Switzerland.}

\begin{abstract}
Femtosecond pump probe experiments are reported on quasiparticle relaxation
and recombination in YSr$_{x}$Ba$_{2-x}$Cu$_{4}$O$_{8}$ as a function of
temperature and polarization. The data show a 2-component relaxation similar
to YBa$_{2}$Cu$_{3}$O$_{7-\delta }$, one component being assocated with the
superconducting transition, and the other with the pseudogap below $T^{\ast
} $. The relaxation time $\tau _{p}$ associated with the pseudogap is found
to be $T$-independent, while the relaxation time $\tau _{g}$ of the
component obseved only below $T_{c}$ exhibits a clear divergence near $T_{c}$%
. A strong polarisation anisotropy of the picosecond transient is observed
below $T_{c}$ which is attributed to the anisotropy of the probe transition
matrix elements.
\end{abstract}

\maketitle

Femtosecond time-domain spectroscopy is capable of giving important
information on the quasi-particle excitations and a low-energy structure of
correlated electron systems, and particularly high-$T_{c}$ superconducting
cuprates (HTSC). Measurements on YBa$_{2}$Cu$_{3}$O$_{7-\delta }$ (Y123) at
different doping levels have shown two-component relaxation dynamics on the
picosecond timescale, that was attributed to the simultaneous existence of
two gaps in optimally doped and overdoped regions \cite{Demsar}. Moreover,
similar two-component relaxation dynamics has been thus far observed in
several HTSC \cite{SmithTa,GuyBiSCO,Hg1223,Eesley,Han,Kaindl,Reitze},
suggesting this being a general feature in these materials. The magnitude
and sign of the two components at temperatures below $T_{c}$ were found to
depend on the material, probe wavelength \cite{Stevens}, and also the
effects of probe polarization dependence were investigated on untwinned Y123
single-crystals \cite{Guy}, where a response parallel and perpendicular to
the Cu-O chains was separately probed and found to be different. In some
cases (Tl2201\cite{SmithTa}, Bi2212 \cite{GuyBiSCO})\ the signs of the
different components observed in the relaxation were opposite. However, the
anisotropy of the photoinduced signal with respect to the probe pulse
polarization has not been discussed in detail thus far.

The purpose of this paper is two fold. First, we show that the 2-component
relaxation behavior observed in Y123\ is similar also in YBa$_{2}$Cu$_{4}$O$%
_{8}$ (Y124), and second, utilizing the fact that Y124 has a fixed oxygen
content and a well-defined untwinned orthorhombic structure, to probe the
polarization dependence of the photoinduced signal.

We report on a systematic investigation of Y124 \ and YBa$_{1.6}$Sr$_{0.4}$Cu%
$_{4}$O$_{8}$ (Y124:Sr) using the usual femtosecond time-resolved pump-probe
technique. As discussed in detail previously \cite{Mihailovic}, a short ($%
\sim $80fs) pump laser pulse excites the carriers in the sample.
Photoexcited electrons and holes with energies on the order of photon energy
quickly thermalize via electron-electron and electron-phonon thermalization,
reaching states just above the gap in a time short compared to the pulse
duration. The gap in the density of states presents a relaxation bottleneck,
and the relaxation of photoexcited carrier density near E$_{F}$ is measured
through measurement of small changes in the optical reflectivity $\Delta R/R$
or transmittance $\Delta T/T$ of the sample as a function of a time-delay
between the pump and probe pulses. In these experiments, a Ti:sapphire
mode-locked laser, which operated at \ a 78 MHz\ repetition rate, was used
as a source of both pump and probe light pulses. The wavelength of the
pulses was centered at approximately $\lambda $ $\approx $ 800nm (1.58eV)
and the intensity ratio of pump and probe pulses was approximately 100:1.
The pump and probe beams were crossed on the sample's surface, where the
angle of incidence of both beams was less than 10$^{o}$. The diameter\ of \
both beams on the surface was $\sim $100$\mu $m and the surface was parallel
to a ab-crystal plane. The typical energy of pump pulses was 0.2nJ (1.25$%
\times 10^{9}$eV), which produces a weak perturbation of the electronic
system with approximately 3$\times 10^{10}$ thermalized photoexcited
carriers per pulse. (The approximation is based on the assumption that each
photon creates $\hbar \omega /\Delta $ thermalized photoexcited carriers,
where $\Delta \approx 40$meV is of the order of the superconducting gap.)
The train of the pump pulses was modulated at 200kHz \ with an acousto-optic
modulator and the small optical changes were resolved out of noise with the
aid of phase-sensitive detection using EG$\&$G digital lock-in amplifier
model 7265. The pump and probe beams were also cross-polarized to reduce
scattering of pump beam into the detector (avalanche photodiode).\ A
detailed description of the experimental technique and the theory of
excitation and relaxation of the photoexcited carriers in superconductors
with different gap structures can be found in refs. \cite{Mihailovic,Kabanov}%
.

The two Y124 samples used for this investigation were prepared in Z\"{u}rich
by a nonstoichiometric flux-growth technique, using a BaO-CuO eutectic
mixture as the flux. In the Sr-doped compound, part of the Ba was
substituted with Sr. The crystals grew at pressure of 900bar and \
temperatures, which ranged from 1000$^{0}$C to 1120$^{0}$C. So obtained
crystals had a thickness of approximately 100$\mu $m in c-axis direction and
the dimensions of 0.4$\times $0.2 mm$^{2}.$ The crystal axis were determined
by x-ray analysis. Details of the growth method and the techniques used for
characterization of the samples are given in ref. \cite{Karpinski}.\ 

The photoinduced (PI) reflection $\Delta R/R$ as a function of time at
different temperatures is shown in Fig. 1 for the direction of polarization
of probe pulse along $a$-axis and $b$-axis of Y124. The chains are parallel
to the $b$-axis and the PI response in this directions clearly shows a
presence of at least two relaxation process with different sign of $\Delta
R/R.$ At temperatures $T>T_{c}$ we see a signal with a positive $\Delta R/R$
and with a relatively fast relaxation time ($\tau _{P}\simeq $0.2ps). As the
temperature is lowered below T$_{c},$ a second component with a longer
relaxation time ($\tau _{G}\approx $2ps) and negative sign starts to appear.
This kind of behavior of the two components is also present in PI response
with the probe polarization along $a$ axis, which can be clearly seen, if
data is presented on a logarithmic scale as in the insert of Fig 1.\ The
relaxation of $\Delta R/R$ after 200fs can be modeled with a function of the
form $\Delta R($t,T$)/R=G($T$)\exp (-$t/$\tau _{G})+P($T$)\exp (-$t/$\tau
_{P})$, where the temperature dependant amplitudes $G($T$)$ ($G($T$)$ $=0$
for $T\geq T_{c}$) and $P($T$)$ can have the same or opposite sign below T$%
_{c}$, depending on the polarization direction of the probe pulse. Fits,
which are presented by a continuous line in Fig. 1, are in a good agreement
with the data. We note that above T$_{c}$ the data shows a slight departure
from a simple exponential decay. Similar behavior above T$_{c}$ has also
been reported in ref. \cite{Eesley}, where two components of opposite sign
below T$_{c}$ and one component slightly non-exponential above T$_{c}$ were
observed on Tl$_{2}$Ba$_{2}$Ca$_{2}$Cu$_{3}$O$_{10}$. Whether a stretch
exponential decay is relevant or the presence of an additional component is
the reason for the discrepancy, we cannot assert with a good degree of
certainty from the present data.

In addition to the two picosecond components, a weak oscillatory component
with frequency of 3.0($\pm $0.3)THz can also be seen above T$_{c}$. This is
shown in Fig. 1 c). Such an oscillatory component of the same frequency was
already observed with pump-probe measurements on Y124 at room temperature 
\cite{Misochko}.

We analyze the temperature dependences of relaxation time $\tau _{G}$ and of
both amplitudes $G($T$)$, $P($T$)$ in the same way as was done previously
for Y123 \cite{Demsar}. The temperature dependance of quasi-particle
recombination time in a superconductor with a temperature dependant gap $%
\Delta $(T) below T$_{c}$ is described by equation \cite{Kabanov} 
\begin{equation}
\tau _{G}=\frac{\hbar \omega _{ph}^{2}\ln \{(\frac{E_{l}}{2N(0)[\Delta
(0)]^{2}}+e^{-\Delta (T)/k_{B}T})^{-1}\}}{12\Gamma _{\omega _{ph}}[\Delta
(T)]^{2}},  \label{relaxtime}
\end{equation}
where $\omega _{ph}$ is a typical phonon frequency, $N(0)$ is the density of
states (DOS), $\ \Gamma _{\omega }$ is a characteristic phonon linewidth and 
$\Delta (0)$ \ is the value of the gap at T=0 K. In Fig. 2 we show divergent
behavior near T$_{c}$ in the temperature dependance of $\tau _{G}$, \
predicted by the Eq. (\ref{relaxtime}) (solid curve). \ The data points were
obtained from the fits of the time evolution of $\Delta R/R$ for Y124 along
the $a$ and $b$ axes and for Y124:Sr along the $b$ axis. In the fits we have
used the following values of parameters in Eq. (\ref{relaxtime}) $\omega
_{ph}=400$ cm$^{-1}$,\ \ $\Gamma _{\omega }$=10 cm$^{-1}$, \ $N(0)$\ =5 eV$%
^{-1}$cell$^{-1}$spin$^{-1}$ and a BCS functional form for the temperature
dependant gap $\Delta (T)$. The fitting parameter was $\Delta (0).$

In Figure 3 we compare the measured temperature dependence of the signal
amplitudes $G($T$)$, $P($T$)$ with those predicted by the theoretical
expressions\cite{Kabanov} for BCS temperature dependant gap 
\begin{equation}
G(\text{T})=\frac{E_{l}/[\Delta (T)+k_{B}T/2]}{1+\frac{2\nu }{N(0)\hbar
\Omega _{c}}\sqrt{2k_{B}T/\pi \Delta (T)}\exp [-\Delta (T)/k_{B}T]}
\label{Tdepgap}
\end{equation}
and for temperature independent gap 
\begin{equation}
P(\text{T})=\frac{E_{l}/\Delta _{p}}{1+\frac{2\nu }{N(0)\hbar \Omega _{c}}%
\exp [-\Delta _{P}/k_{B}T]}.  \label{Tindepgap}
\end{equation}
In the fits we used $\nu =$18 for the number of modes interacting with
quasi-particles, $\Omega _{c}=$0.1eV for a typical phonon cutoff frequency
and $N(0)$\ =5 eV$^{-1}$cell$^{-1}$spin$^{-1}$ for the DOS. The agreement
between the data and the theory is seen to be very good.

In Fig 3 c) we show a polarization dependence of both amplitudes $G(T,$ ${%
\theta })$, $P(T,$ ${\theta })$ for Y124 at $T=45$ K. $P($T$)$ does not show
any polarization dependance, which was also confirmed by polarization
measurements above T$_{c.}$ In contrast, the polarization dependance of $G($T%
$)$ shows a significant angular dependence, exhibiting a change of sign as
the polarization direction is changed from parallel to perpendicular to the
Cu-O chains.

This angular dependance can be understood in terms of the anisotropy of the
probe transition matrix elements. We first note that in spite of the fact
that Y124 is orthorhombic, the main contribution to the probe signal comes
from interband resonance transitions \cite{Stevens} which involves atomic
wave functions in which the dipolar matrix elements have a 4-fold rotational
symmetry around the $c$ axis. Without specifying exactly which transitions
are involved, we can write the absorption coefficient in terms of the Fermi
golden rule: 
\[
\alpha \propto \int d\epsilon N(\epsilon )N(\epsilon +\omega )|\mathbf{M}%
(\epsilon ,\omega )|^{2}f(\epsilon )(1-f(\epsilon +\omega )) 
\]
Here $N(\epsilon )$ is the density of electronic states, $f(\epsilon )$ is
the distribution function for holes, $\hbar \omega $ is the energy of the
probe photons and $\mathbf{M}(\epsilon ,\omega )$ is the dipole matrix
element for the transition. For small perturbations, $\Delta R$ is
proportional to the photoinduced absorption $\Delta \alpha $, so: 
\begin{equation}
\Delta R\propto \Delta \alpha \propto \int d\epsilon N(\epsilon )|\mathbf{M}%
(\epsilon ,\omega )|^{2}(f^{^{\prime }}(\epsilon )-f(\epsilon ))
\label{apsorchange}
\end{equation}
where $f^{^{\prime }}(\epsilon )$ is the nonequilibrium distribution
function of the charge carriers. The integral is taken in the vicinity of
the Fermi energy $E_{F}$ over the width of the resonance \cite{Stevens} and
we assumed for simplicity that $N(\epsilon +\Omega )$ is constant within the
resonance width.

If $\mathbf{M(\epsilon ,\omega )}$ is constant over the whole range of
energies $-\hbar \omega <\epsilon -E_{F}<\hbar \omega $, then $\Delta
R=\Delta \alpha =0$ because of the conservation of particles$.$ In other
words, the photoinduced \emph{increase} in absorption due to probe
transitions $\mathsf{1}$ (see Figure 4) originating from the photoexcited
electron states at $\epsilon -E_{F}\gtrsim \Delta $ just above the gap to
unoccupied hole states at $\epsilon -E_{F}\sim \hbar \omega +\Delta $
exactly cancel the \emph{decrease} in absorption due to transitions $\mathsf{%
2}$ originating from occupied electronic states at $\epsilon -E_{F}\lesssim
-\Delta $ just below the gap to unoccupied states at $\epsilon -E_{F}\sim
\hbar \omega -\Delta $. The same is true for the hole transitions \textsf{3}
and \textsf{4} (see Fig. 4).

Turning to the situation in hand, we assume that $\mathbf{M(\epsilon ,\omega
)}$ is not constant and\ proceed to derive an expression for the
polarization anisotropy of the probe absorption. The general expression for
the square of the dipolar matrix element can be written as: 
\[
|\mathbf{M}(\epsilon )|^{2}=M_{x}^{2}(\epsilon )\sin ^{2}{(\theta )}%
+M_{y}^{2}(\epsilon )\cos ^{2}{(\theta )} 
\]
where $\theta $ is the angle between polarization of light and the $b$ axis.
For an orthorhombic structure $M_{x}\neq M_{y}$. On the other hand, the main
contribution to $M$ which comes from atomic wave functions is independent of 
$\theta ,$ so we can expand $M$ in the vicinity of the Fermi energy and
express matrix element in the form: 
\begin{equation}
M_{x,y}(\epsilon )=M_{0}+\gamma _{x,y}\epsilon  \label{matrix}
\end{equation}
where $\gamma _{x,y}=dM_{x,y}/d\epsilon $ and the derivative is taken at the
Fermi energy. Substituting this expression into Eq.(\ref{apsorchange}) we
obtain a qualitative description of the angular dependence of the probe
signal $G({\theta })$ and $P({\theta })$: 
\begin{equation}
\Delta R\propto \Delta \alpha \propto M_{0}(\gamma _{x}\sin ^{2}{(\theta )}%
+\gamma _{y}\cos ^{2}{(\theta )})\Delta n  \label{apsorchange2}
\end{equation}
Here $\Delta n$ is the number of photoexcited quasiparticles, as described
in the Ref.\cite{Kabanov}. The values of $\gamma _{x,y}$ can be doping
dependent and can easily have different sign, depending strongly on $\omega $
and the material's band structure in the vicinity of the resonance $\epsilon
-E_{F}\sim \hbar \omega \pm \Delta $. However, since the probe polarization
anisotropy is a consequence of the anisotropy of the probe transition matrix
elements, Eq. (\ref{apsorchange2}) unfortunately does not give any direct
information regarding the anisotropy of the low-energy electronic gap
structure. In Fig. 3 c) we plot the polarization dependence of photoinduced
reflectivity amplitude given by Eq. (\ref{apsorchange2}) using $\gamma
_{x}/\gamma _{y}=35/(-10)$, which is shown to describe the main features of
the data quite well, despite the fact that Eq.(\ref{apsorchange2}) is based
on the oversimplified expansion of the matrix elements (Eq. \ref{matrix}) -
neglecting higher order terms.

An alternative model discussing carrier relaxation dynamics in HTSC has been
recently proposed \cite{SegreHan}, whereby the main contribution to $\Delta
R/R$ in the region at $\omega \sim 1.5$ eV comes from spectral weight
transfer in the real part of the optical conductivity $\sigma (\omega )$
from $\omega =0$ to $\omega \approx \Delta $ . Since $\Delta \sigma $ is
always positive this leads to negative changes in the real part of the
dielectric function, independent of polarization. The change of sign in $%
\Delta R/R$ as a function of probe polarization which we observe in Y124 is
clearly inconsistent with the model proposed in ref. \cite{SegreHan}.

In conclusion, the femtosecond relaxation dynamics in Y124 is found to be
very similar to that reported previously in Y123 \cite{Demsar}. The results
are also consistent with time-resolved terahertz spectroscopy measurements 
\cite{Averitt}, which directly probed the recovery of the condensate after
photoexcitation, both agreeing well with the model \cite{Kabanov}. The
observed probe polarization dependence and sign change of the transient
signal below T$_{c}$ is described well by a model which considers the
anisotropy of the probe transition matrix elements in Y124, and is not
directly related to the symmetry of the order parameter.

\bigskip 

The authors would also like to acknowledge A.Mironov of the Chemical
Department Moscow State University for single crystal x-ray investigation of
the samples. 

\section{Figure captions \ }

Figure 1: \ The photoinduced reflection $\Delta R/R$ from Y$_{1}$Ba$_{2}$Cu$%
_{4}$O$_{8}$ at different temperatures above and below T$_{c}$ as a function
of time, measured a) with the polarization of probe pulse in the direction
parallel to the crystal axis a \ and b) parallel to (the direction of
chains) the crystal axis b. The insert shows the data at 28K and 74K already
presented in a) on the logarithmic scale, so that two component decay can be
easily observed. \ In c) we show weak oscillations of $\Delta R/R$ with
frequency of 3.0($\pm $0.3)THz observed above T$_{c}$ and attributed to a
coherent phonon mode.

Figure 2: \ The relaxation times $\tau _{G}$ \ as a function of temperature
a) for Y$_{1}$Ba$_{2}$Cu$_{4}$O$_{8}$, with the direction of polarization of
probe pulse along a axis (open triangles) and b axis (squares), and b) for
YBa$_{1.6}$Sr$_{0.4}$Cu$_{4}$O$_{8}$\ with the direction of polarization of
probe pulse along b axis (squares).

Figure 3: a) \ The temperature dependance of photoinduced \ amplitudes $G($T$%
)$ for Y$_{1}$Ba$_{2}$Cu$_{4}$O$_{8}$, measured with the polarization of
probe pulse along a axis (circles) and b axis (open triangles), and for YBa$%
_{1.6}$Sr$_{0.4}$Cu$_{4}$O$_{8}$ with the polarization along b axis
(squares). The values of $\Delta (0)$ obtained from the fits are also shown.
b)\ The photoinduced amplitudes $P($T$)$ for Y$_{1}$Ba$_{2}$Cu$_{4}$O$_{8}$
and YBa$_{1.6}$Sr$_{0.4}$Cu$_{4}$O$_{8}$ as a function of temperature,
measured at the same polarizations as in a). We also report the obtained
values of \ $\Delta _{p}.$ c) The polarization dependance of photoinduced
amplitudes $G($T$=45$ K$,{\theta })$ and $P($T$=45$ K$,{\theta })$ for Y$%
_{1} $Ba$_{2}$Cu$_{4}$O$_{8}$, together with the polarization dependence
given by Eq. \ref{apsorchange2} (solid line).

Figure 4: The four possible transitions, which can give rise to a
photoinduced probe signal arising form a change in quasiparticle population $%
\Delta n.$ If the transition probabilities do not cancel a non-zero
photoinduced reflectivity transient is observed. The polarization anisotropy
of the probe signal arises due to the transitions having different
probabilities in different directions relative to the crystal axes. Note:
the scheme is drawn for T = 0K.\

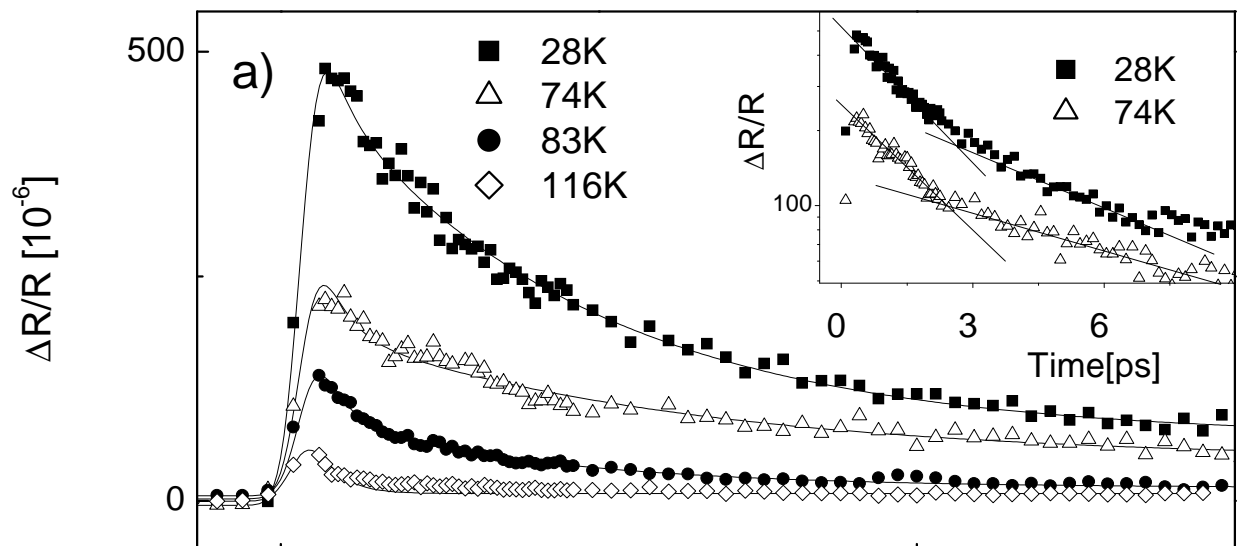
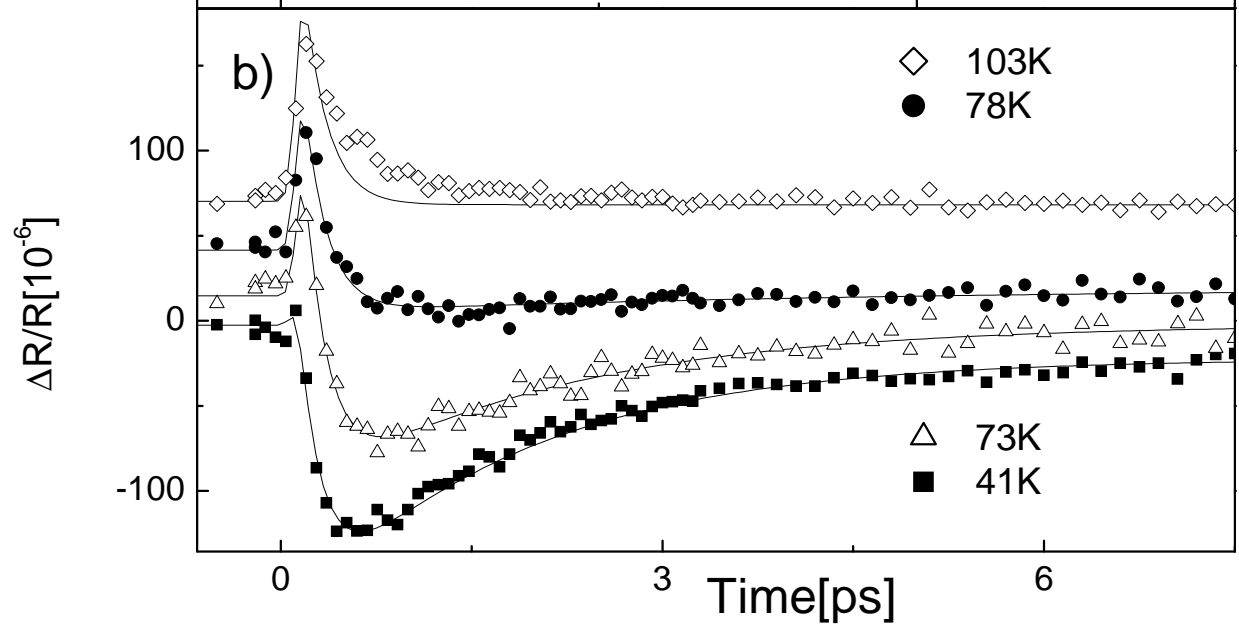
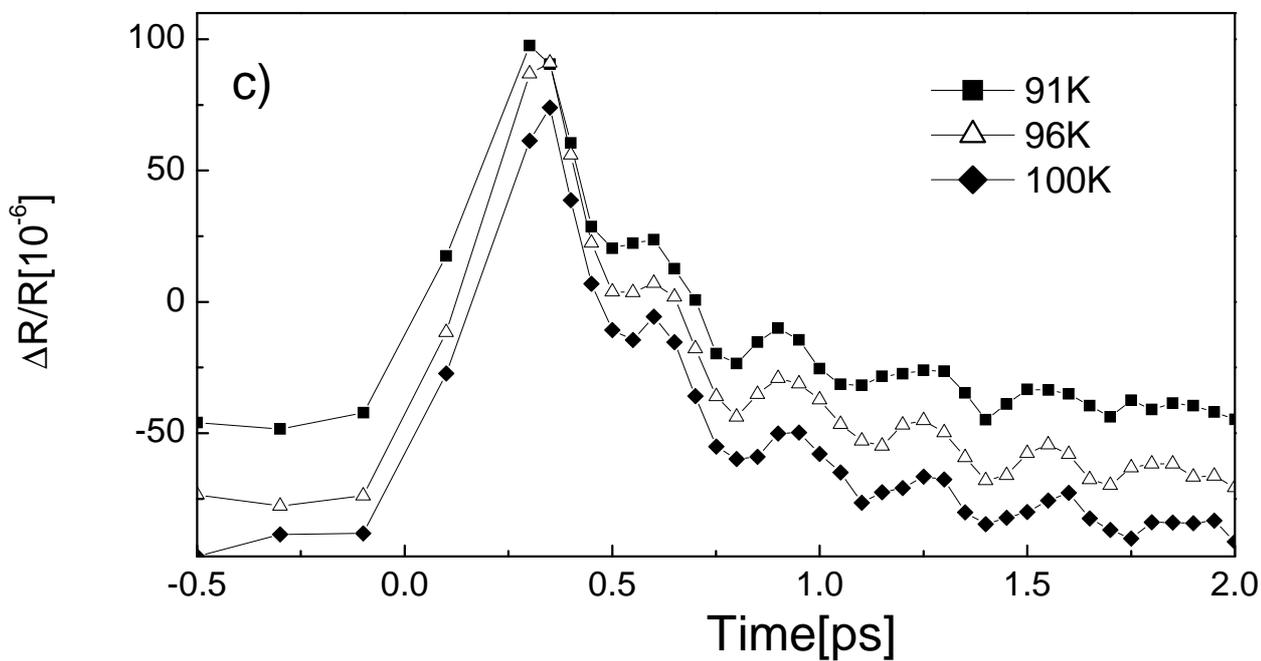
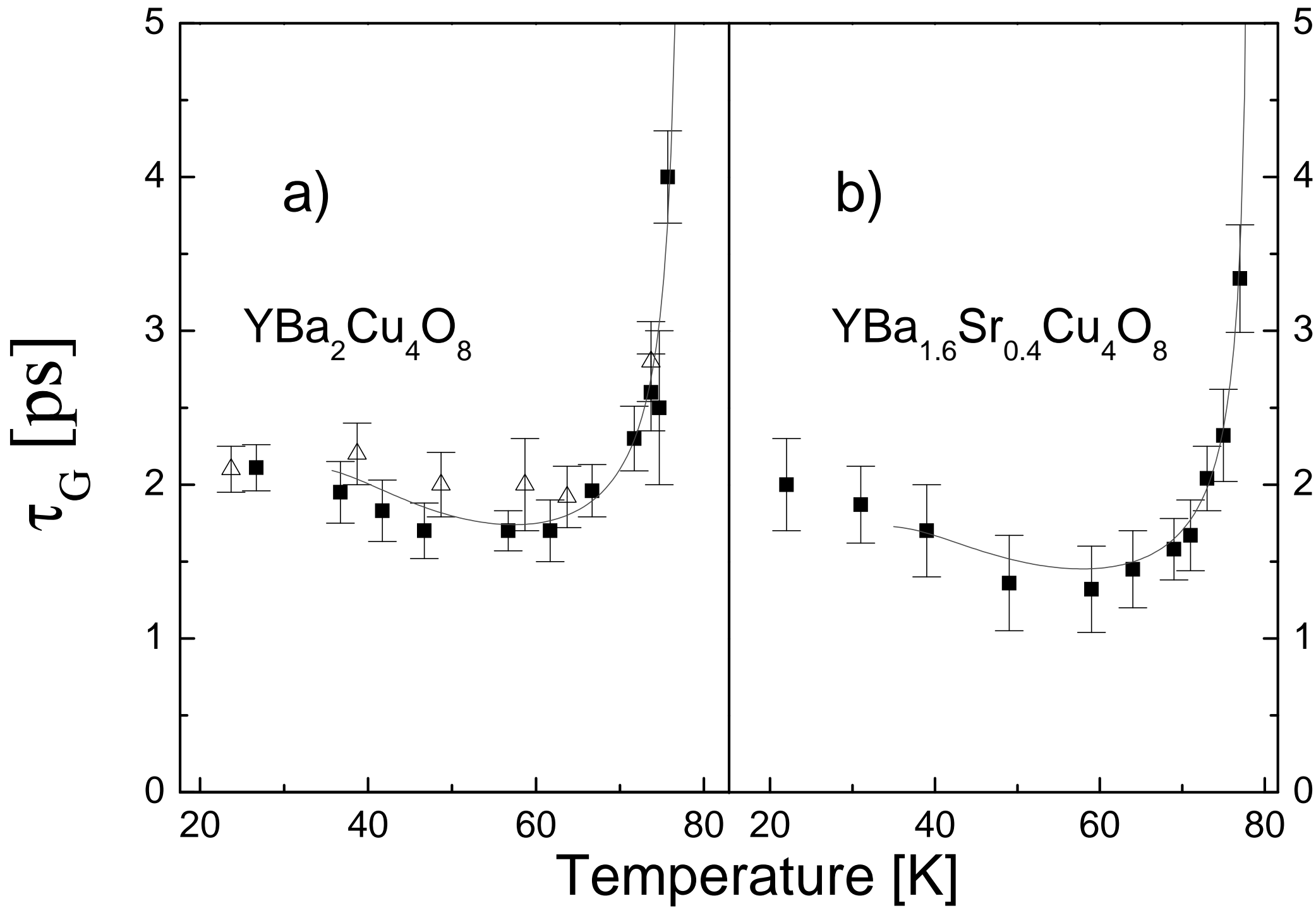
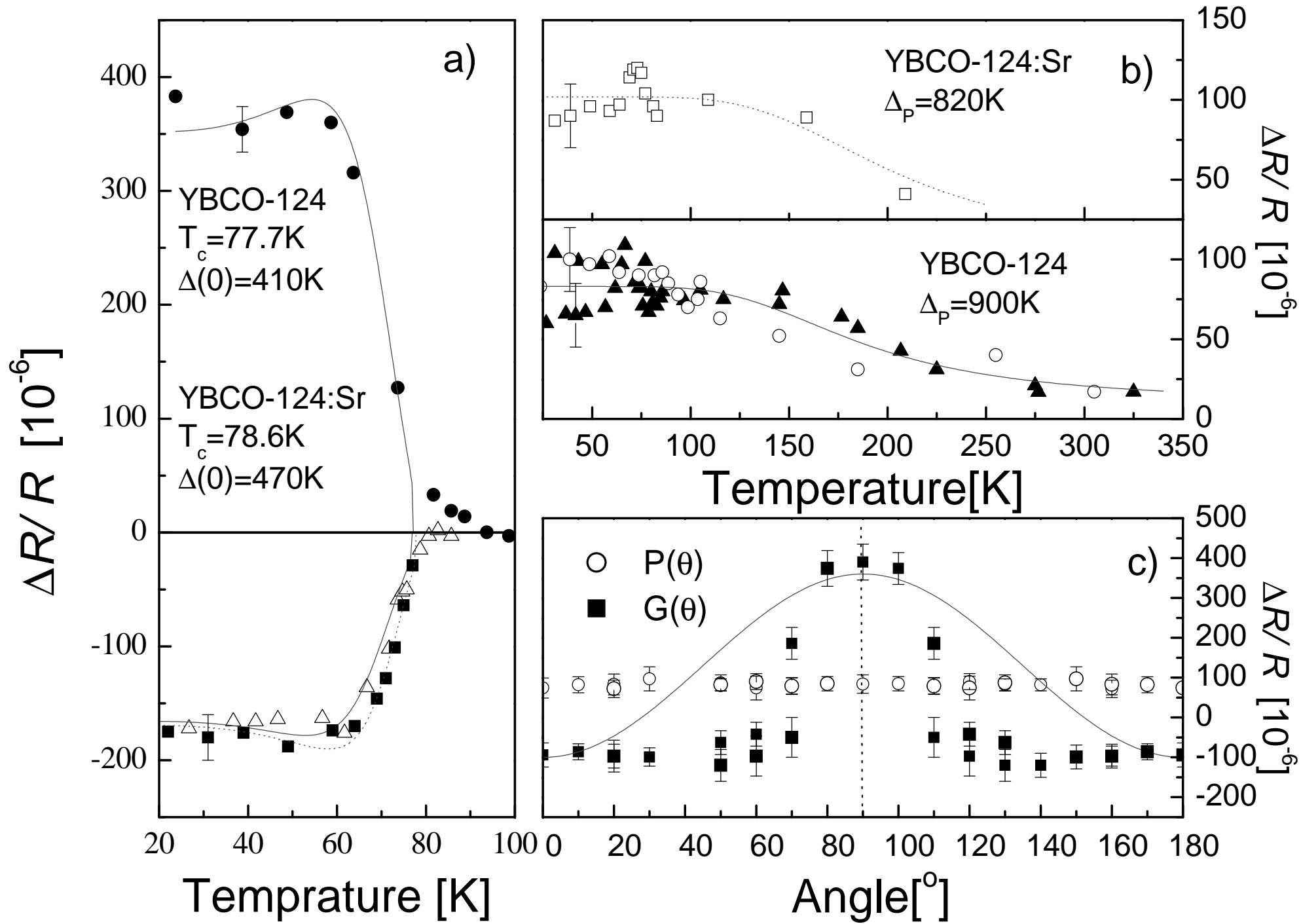
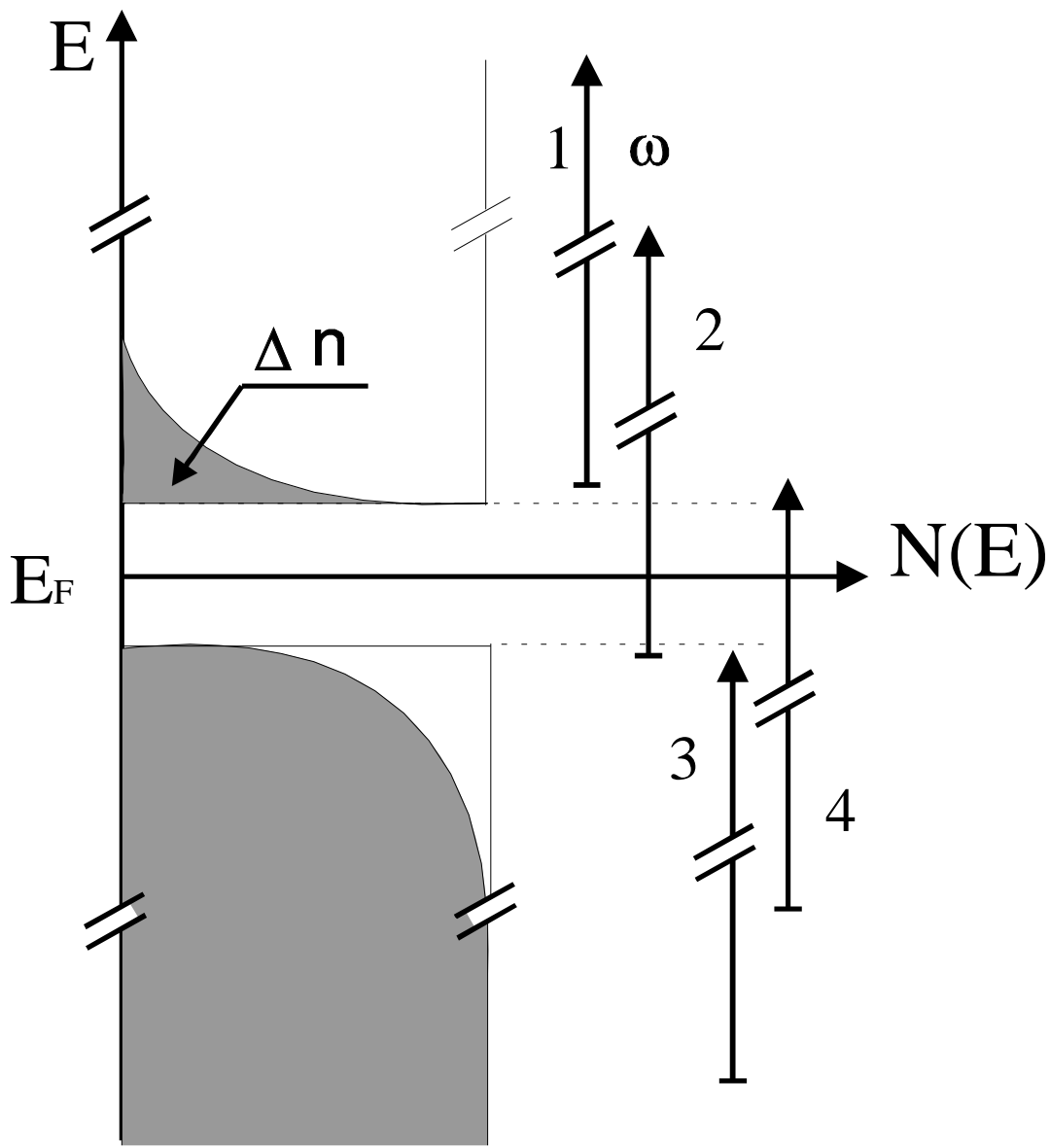
\end{document}